\documentclass{article}

\usepackage[final,nonatbib]{neurips_2018}

\usepackage[utf8]{inputenc} % allow utf-8 input
\usepackage[T1]{fontenc}    % use 8-bit T1 fonts
\usepackage{hyperref}       % hyperlinks
\usepackage{url}            % simple URL typesetting
\usepackage{booktabs}       % professional-quality tables
\usepackage{amsfonts}       % blackboard math symbols
\usepackage{nicefrac}       % compact symbols for 1/2, etc.
\usepackage{microtype}      % microtypography
\usepackage{graphicx}

\title{Statement networks: a power structure narrative as depicted by newspapers}

\author{
  Shoumik Sharar Chowdhury \\
  KolpoKoushol\\
  Dhaka, Bangladesh\\
      \texttt{shoumikchow@gmail.com} \\
  \And
  Nazmus Saquib\\
  MIT Media Lab\\
  Cambridge, MA 02139 \\
  \texttt{saquib@mit.edu} \\
  \And 
  Niamat Zawad\\
  KolpoKoushol\\
  Dhaka, Bangladesh\\
  \texttt{zawad1879@gmail.com} \\
   \And 
  Manash Kumar Mandal\\
  KolpoKoushol\\
  Dhaka, Bangladesh\\
  \texttt{manashmndl@gmail.com} \\
  \And 
  Syed Haque \\
  KolpoKoushol\\
  Dhaka, Bangladesh\\
  \texttt{syedarehaq@gmail.com}\\
}

\begin{document}

\maketitle

\begin{abstract}
    We report a data mining pipeline and subsequent analysis to understand the core periphery power structure created in three national newspapers in Bangladesh, as depicted by statements made by people appearing in news. Statements made by one actor about another actor can be considered a form of public conversation. Named entity recognition techniques can be used to create a temporal actor network from such conversations, which shows some unique structure, and reveals much room for improvement in news reporting and also the top actors' conversation preferences. Our results indicate there is a presence of cliquishness between powerful political leaders when it comes to their appearance in news. We also show how these cohesive cores form through the news articles, and how, over a decade, news cycles change the actors belonging in these groups. 
    
\end{abstract}

\section{Introduction}

Machine Learning techniques like named entity recognition opened up many avenues of novel analysis on social media and newspapers. In this work, we take a different turn by using named entity recognition to identify highly mentioned actors in newspapers and create a network of them by parsing the statements they make about one another. Studying some properties of this unique network reveals some interesting characteristics that could be attributed to both the journalistic style of different newspapers and the culture of political statements in Bangladesh. 

Previous works on newspaper text analysis looked at how a news outlet is biased towards particular political ideologies \cite{groseclose2005measure}\cite{akhavan1998framing}, or comparison between the writing style of different news outlets in different countries \cite{kirilenko2012computer}. Additionally, there has been work on comparing newspapers in local, regional and national level \cite{snyder1977conflict} and comparing information gathered from surveying people in the locality and related news articles \cite{martin2013network}. In this paper we show that the political leaders appearing in Bangladeshi newspapers have a tendency to create statements about other powerful politicians.

\section{Data acquisition and processing}
\label{gen_inst}

\begin{figure}[h]
\centering
\includegraphics[width=12cm, height=6cm]{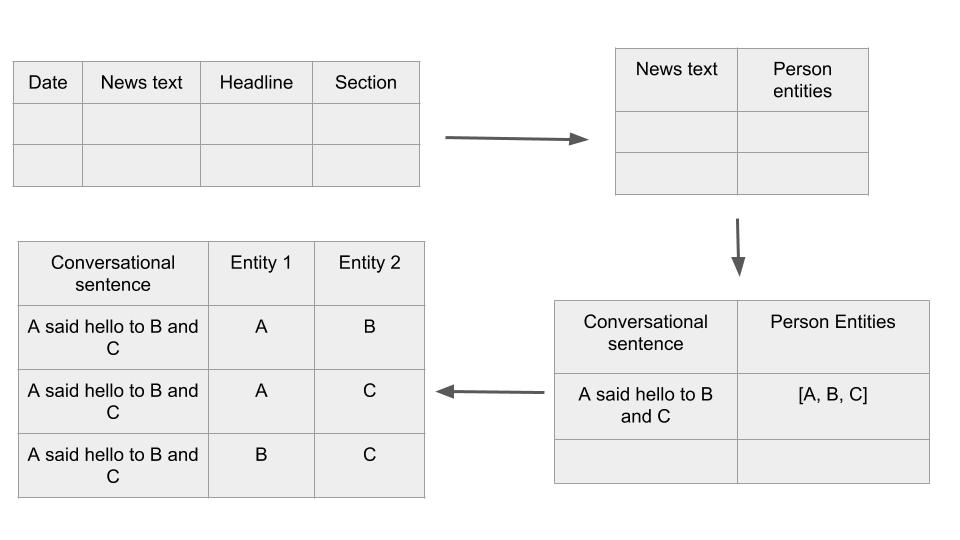}
\caption{Creation of statement network from news article}
\label{fig:statement_network_creation}
\end{figure}

\subsection{Scraping and tagging news}
Three major English dailies, The Daily Star, Dhaka Tribune, and New Age, were scraped for this research. Necessary information, like news text, headline, date of publication, news category, etc. were obtained from all news sections. Information like breadcrumb, subheadings, taglines that were not consistently available across all three newspapers were ignored. The Daily Star data was obtained in the time range 2008-01-01 to 2017-12-01 (119 months), Dhaka Tribune data was obtained in the time range 2013-04-06 to 2016-06-14 (38 months), and New Age data was obtained in the time range 2016-10-01 to 2018-04-22 (18 months). Even though the time range wasn't consistent across the newspapers, they could be compared because the volume of data from each newspaper was sufficient to form a large enough statement network and study the dynamics of network evolution.

\subsection{NER and network formation}
After scraping the three newspapers, a named-entity recognition (NER) tagger \cite{finkel2005incorporating} was used to find all the person labels that appear in the news text. In some cases, some names or entities appeared to be broken up into two separate names. Some obvious ones that appeared frequently were fixed by concatenating the names. For example, if Barack and Obama were identified as two different people, they were considered as Barack Obama to maintain consistency and avoid repetition in the network.

Next, a statement network was created for each newspaper. In order to create it, each news text was searched for quotation marks and words like `said', `asked', `told', `spoke', `speak', `says', `added', `declare' and `alleged' (and their inflected forms) to look for sentences that indicated a speech or a conversation. These sentences were chosen only if two or more entities tagged previously appeared in them. On random manual checking of the sentences, this method proved to be largely effective to select conversations among entities or about entities. An indirect statement network was then created among these entities by finding all possible combinations of entities that are identified as a person. For example, if entity \textit{A}, \textit{B} and \textit{C} appeared together in one sentence, the combinations would be \textit{AB}, \textit{BC} and \textit{AC}. The process of the creation of statement network is illustrated in \ref{fig:statement_network_creation}.

\section{Analysis of statement network and its dynamics}
\label{others}

\subsection{Core-periphery structure of news statement network}

To investigate the power structure in a statement network, we compute its core-periphery hierarchy. Often time an investigation of the core-periphery structure reveals the tight cliques between people of elite status \cite{borgatti2000models} \cite{alba1978elite}. To identify the core-periphery structure in a network we use the k-core algorithm \cite{wasserman1994social}. To be included in k-core, a node has to be connected to at least k other nodes in the group. This measure helps identify the tightly connected groups within the network. 

\begin{figure}[h]
  \centering
  \includegraphics[width=\linewidth]{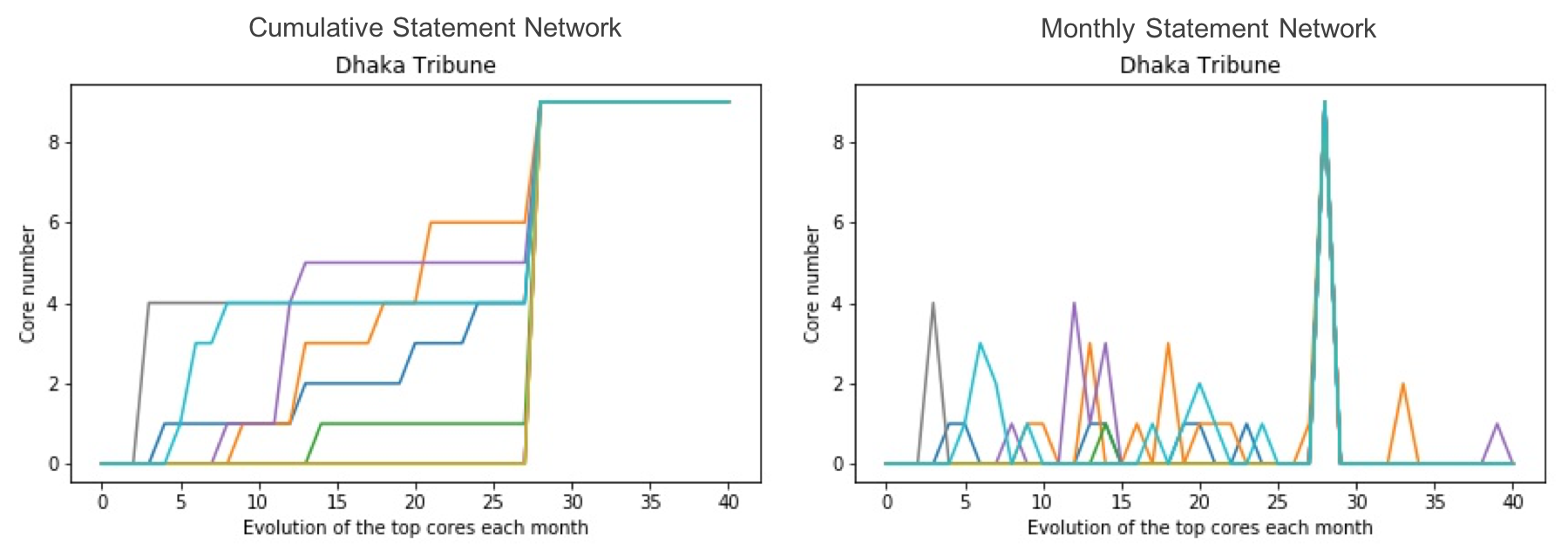}
  \caption{Evolution of the members in top core in Dhaka Tribune}
  \label{fig:evolution_top_core}
\end{figure}

The investigation of top cores revealed an interesting picture. After manual inspection of the actors appearing in the topmost cores, we identified them as political elite people such as the Prime Minister, opposition leader, ministers, top members of different political parties, etc. The number of cores are 9, 35 and 39 respectively for Dhaka Tribune, New Age, and The Daily Star.

This led us to investigate of how such a top core was formed and how it evolved over time. Figure \ref{fig:evolution_top_core} shows an example of a top core's evolution in Dhaka Tribune. There are nine people in the top core (the 9 curves in figure \ref{fig:evolution_top_core}). We track how their core rank changes over 38 months. A cumulative statement network for a month x was formed by adding all the edges that appeared up to month x. On the 38th month, we get an accumulated network in this way (that we used to calculate the core periphery hierarchy above). Additionally, we investigated the monthly core periphery structure by accumulating the edges created from daily news over one-month periods.

In the cumulative network, we noticed a sudden phase transition at the 27th month that forms the top core. We noticed that these people tend to frequently appear together in different news and ultimately fall in the same core because of such a transition. The monthly network's core ranks reveal that they all appeared together in a few news on the 27th month. Thus the cumulative network retains all those edges and keeps them in the top core until the end.

\subsection{Dynamics of news cycle}

\begin{figure}[h]
  \centering
  \includegraphics[width=\linewidth]{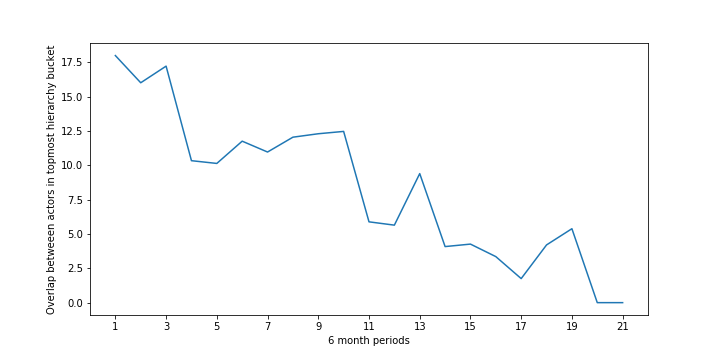}
  \caption{Evolution of overlap between actors in hierarchy buckets}
  \label{fig:hierarchy_bucket_overlap}
\end{figure}

The topics or persons that are covered in newspapers \cite{bucy2007media} or internet \cite{leskovec2009meme} change over time, which is generally known as news cycle. To understand the effect of news cycle on the top most cores, we created separate statement networks in The Daily Star for every 6 month period. We created a top most hierarchy bucket by collecting the nodes with core-number greater than 7 for each of these statement networks. Then we tracked the membership of these topmost hierarchy buckets over 21 six-month periods (10 years).

In Figure \ref{fig:hierarchy_bucket_overlap} we saw that the overlap, which was calculated as a percentage overlap between the topmost hierarchy bucket on 6 month intervals, between these buckets was not significantly high and it actually even diminishes over time. This shows that over a long period of time, as the news cycles change, the actors in the topmost cores change. So the cohesiveness of the top cores are affected by news cycle.

\subsection{Newspapers' conversational style}
\begin{figure}[h]
  \centering
  \includegraphics[width=\linewidth]{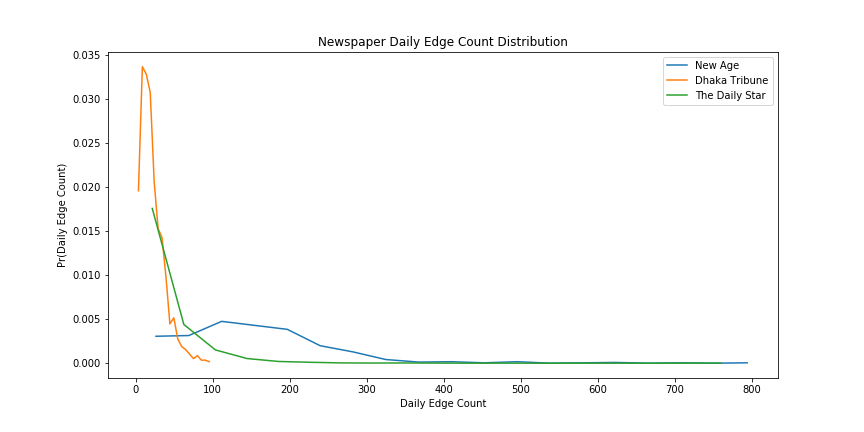}
  \caption{Daily edge count distribution for the newspapers}
  \label{fig:conversational_style}
\end{figure}

In Figure \ref{fig:conversational_style} we present the daily edge count distribution of the statement networks extracted from the newspapers. An edge was created in the network if an article connects two nodes (persons) through a conversation like sentence. Higher edge count on a particular day would indicate that the newspaper covered more conversational news on that day. From the daily edge count distribution we notice the difference between the conversational styles of the newspapers. Dhaka Tribune has a lot more days in the timeline where they have a lot less conversational articles (i.e., less number of statements made). In comparison to that, New Age has more conversations as the daily edge distribution starts from a lower value and flattens out soon. Both The Daily Star and New Age have long tails which show that on some days they have more than average conversational news published.

\section{Discussion}

A lack of statements made about general topics (such as development issues, governmental organizations etc.) is revealed in our investigation, as the most interconnected cores do not contain any other personalities other than top political leaders from government and opposition parties. This could be an artifact of how journalists cover an event (conversational style) in different newspapers, or a culture of political debates that do not usually take account of pressing issues. Causal relationships cannot be derived from the data that we have, but this work invites further development of machine learning pipelines (such as stance detection, Bangla named entity recognition for native language newspapers etc.) to investigate statement networks and public statements made by political leaders in Bangladesh.

\medskip

\small

{%\footnotesize
\bibliographystyle{acm}
\bibliography{bib}
}

\end{document}